\documentclass[namedreferences]{SSRv}

\def\etal{\textit{et al }}
\def\d{\mathrm{d}}     
\def\g{\mathrm{g}}     
\def\s{\mathrm{s}}     
\def\S{\mathrm{Sun}}     %
\def\bv{\textbf{v}}     
\def\Hz{\mathrm{Hz}}   

\begin{document}
\begin{article}
\begin{opening}

\title{Testing General Relativity with Atomic Clocks}

\author{S. \surname{REYNAUD}$^1$,
      C. \surname{SALOMON}$^2$,
        P. \surname{WOLF}$^3$ }

\runningauthor{Reynaud, Salomon and Wolf}
\runningtitle{Testing general relativity with atomic clocks}

\institute{
$^1$ Laboratoire Kastler Brossel, ENS, UPMC, CNRS \email{serge.reynaud@upmc.fr}\\
Case 74, Campus Jussieu, 75252 Paris, France\\
$^2$ Laboratoire Kastler Brossel, ENS, UPMC, CNRS \\
D\'epartement de Physique de l'Ecole Normale Sup\'erieure, 75231 Paris, France\\
$^3$ LNE-SYRTE, Observatoire de Paris, CNRS, UPMC\\
61, avenue de l'Observatoire, 75014 Paris, France}

\date{Received; accepted }

\begin{abstract}
We discuss perspectives for new tests of general relativity which are based
on recent technological developments as well as new ideas.
We focus our attention on tests performed with atomic clocks and
do not repeat arguments present in the other contributions
to the present book~\cite{ThisIssue}.
In particular, we present the scientific motivations of
the space projects ACES~\cite{Salomon01} and SAGAS~\cite{Wolf09}.
\end{abstract}
\end{opening}

 \vspace{-0,5cm}

{\small

\section{Introduction}
\label{intro}
Tests of gravity performed in the solar system show a good agreement
with general relativity.
The latter is however challenged by observations at larger
galactic and cosmic scales which are presently taken care through
the introduction of ``dark matter'' or ``dark energy''.
As long as these components are neither detected through non gravitational
means, nor explained as resulting from new physical phenomena,
it remains of the uttermost importance to test general relativity
at all experimentally accessible scales.

In this paper, we recall the basic elements of general relativity
and briefly review the experimental evidences supporting it.
We then discuss perspectives for new tests which are based
on recent technological developments as well as new ideas.
We focus our attention on tests performed with atomic clocks and
do not repeat arguments present in the other contributions
to the present book~\cite{ThisIssue}.

\section{Tests of general relativity}
\label{basic}
General relativity (GR) is built up on two basic ideas which have to be
distinguished.
The first one is the metrical (geometrical) interpretation of gravitation,
which was proposed by Einstein soon after his pioneering work
on special relativity~\cite{Einstein07,Einstein11}.
This identification of gravitation field with the space-time metric is the
very core of GR, but it is not sufficient to fix the latter theory.
In order to select GR out of the variety of metric theories of gravitation,
it is necessary to fix the relation between the geometry of space-time and its
matter content. In GR, this relation is given by the Einstein-Hilbert equation
which was written only in 1915~\cite{Einstein15,Einstein16,Hilbert}.

The metrical interpretation of gravity, often coined under
the generic name of the ``equivalence principle'', is one of the most
accurately verified properties of nature~\cite{Will01}.
Freely falling test masses follow the geodesics of the Riemannian space-time,
that is also the curves which extremize the integral
\begin{eqnarray}
\label{Deltas}
\Delta s \equiv \int \d s \quad,\quad \d s^2\equiv g_{\mu\nu}\d x^\mu \d x^\nu
\end{eqnarray}
$g_{\mu\nu}$ is the metric tensor characterizing the space-time
and $\d x^\mu$ the displacements in this space-time.
As free motions obey a geometrical definition,
they are independent of the compositions of the test masses.
The potential violations of this ``universality of free fall'' property
are parametrized by a relative difference
in the accelerations undergone by two test bodies
in free fall from the same location with the same velocity.
Modern experiments constrain this parameter to stay below
$10^{-12}$, this accuracy being attained in laboratory
experiments~\cite{Adelberger03,Schlamminger08} as well as in space tests using lunar laser
ranging~\cite{Williams96,Williams04} or planetary probe tracking~\cite{Anderson96}.
These results do not preclude the possibility of small violations of the
equivalence principle and such violations are indeed predicted by unification
models~\cite{Damour02}.
Large improvements of this accuracy are expected in the future thanks to
the existence of dedicated space projects \textsc{Microscope}~\cite{Touboul01}
and, on a longer term, \textsc{Step}~\cite{Mester01}.

As another consequence of the geometrical interpretation of gravity,
ideal atomic clocks operating on different quantum transitions measure the same time,
because it is also a geometrical quantity, namely the
proper time $\Delta s$ integrated along the trajectory (see eq.~\ref{Deltas}).
This ``universality of clock rates'' property has also been verified with
an extreme accuracy.  Its potential variations are measured as
a constancy of relative frequency ratios between different clocks
at a level of the order of $10^{-16}$ per year, recently a few $10^{-17}$ per year%
~\cite{Marion03,Bize03,Fisher04,Peik04,Fortier07,Ashby07,Rosenband08}. 
These results can also be interpreted in terms
of a potential ``variation of fundamental constants''~\cite{Flambaum06}, 
thus opening a window on the ``new physics'' expected
to lie ``beyond the standard model''. 
In this domain also, large improvements of the accuracy can be expected
in the future with the space project \textsc{ACES}%
~\cite{Salomon01} and, on a longer term, with projects
using optical clocks~\cite{Wolf09,Schiller09}.

As already stated, GR is selected out of the large family
of metric theories of gravity by the Einstein-Hilbert equation which
fixes the coupling between curvature on one hand, matter on the other one,
by setting the Einstein curvature tensor $G _{\mu\nu}$
to be proportional to the stress tensor $T _{\mu\nu}$
\begin{eqnarray}
\label{GR_gravitation_law}
&&G_{\mu\nu} \equiv R _{\mu\nu} - {1\over2} g _{\mu\nu} R
= {8\pi G  \over c^4} T _{\mu\nu}
\end{eqnarray}
The coefficient is fixed by the Newtonian limit and
determined by the Newton constant $G$ and the velocity of light $c$.
At this point, it is worth emphasizing that
it is not possible to deduce the relation (\ref{GR_gravitation_law})
only from the geometrical interpretation of gravity.
In other words, GR is one member of the family of metric theories of gravity
which has to be selected out of this family by comparing its predictions
to the results of observations or experiments.
The tests performed in the solar system effectively show a good
agreement with GR, as shown in particular by the
so-called ``parametrized post-Newtonian'' (PPN) approach~\cite{Will01}.

The main idea of the PPN approach can be described by
writing down the solution of GR with a simple model
of the solar system where the gravity sources are reduced to the Sun
treated as a point-like motion-less mass $M$
(for the real solar system, see for example \cite{Petit05}).
Using a specific gauge convention where spatial coordinates
are isotropic, the metric element thus takes the following form
(coordinates are $x^0\equiv ct$, the radius $r$, the colatitude
and azimuth angles $\theta$ and $\varphi$)
\begin{eqnarray}
\label{isotropic}
&&\d\s^2 = g _{00}  c ^2 \d t ^2 +  g _{rr} \left( \d r^2  +
 r ^2(\d\theta^2 + \sin^2\theta  \d\varphi^2) \right)
\end{eqnarray}
With this simple model, an exact solution can be found for the metric.
It is convenient to write it
as a power series expansion in the reduced Newton potential $\phi$
(the length scale $GM/c^2$ has a value of the order of 1.5km,
with $M$ the mass of the Sun, so that $\phi$ is much smaller
than unity everywhere in the solar system)
\begin{eqnarray}
\label{GR_metric_expansion}
&&g _{00} = 1 + 2 \phi + 2 \phi^2 + \ldots \;,\quad
 g _{rr} = -1 + 2 \phi + \ldots \;, \quad
\phi \equiv -{GM\over rc^2}
\end{eqnarray}

The family of PPN metrics can then be introduced by inserting
a constant $\beta$ in front of $\phi^2$ in $g _{00}$
and a constant $\gamma$ in front of $\phi$ in $g _{rr}$
(with $\beta=\gamma=1$ in GR).
The values of these PPN parameters affect the predicted motions,
and can therefore be confronted to the observations.
Experiments on the propagation of light have led to
more and more stringent bounds on $\vert\gamma-1\vert$,
with the best current results corresponding to
deflection measurements using VLBI astrometry~\cite{Shapiro04} and
Doppler tracking of the Cassini probe during its 2002 solar
occultation~\cite{Bertotti03}.
Meanwhile, analysis of lunar laser ranging data~\cite{Williams04}
have led to bounds on linear superpositions of $\beta$ and $\gamma$,
and then in constraints on $\vert\beta-1\vert$.
The current status of these tests clearly favors GR
as the best description of gravity in the solar system.

Let us emphasize that the common presentation of this status under the form
``general relativity is confirmed by the tests'' is a bit too loose.
The tests discussed so far do not answer by a final ``yes'' answer
to a mere ``yes/no'' question.
They rather select a vicinity of GR
as the best current description of gravity within the family of PPN metrics.
This warning is not a mere precaution, it is rather a pointer to
possible future progress in the domain.
There indeed exist theoretical models (see for example~\cite{Damour02}) which
deviate from GR while staying within the current observational bounds.
Furthermore, as discussed below, extensions of GR do not necessarily belong
to the PPN family.

This last point is in particular emphasized by the so-called ``fifth force'' tests
which are focused on a possible scale-dependent deviation from the gravity
force law~\cite{Fischbach98}.
Their main idea is to check the $r-$dependence of the gravity
potential, that is also of the component $g _{00}$.
Hypothetical modifications of its standard expression, predicted by unification models,
are often parametrized in terms of a Yukawa potential added to the standard $g _{00}$.
This potential depends on two parameters, an amplitude $\alpha$ measured with respect to
Newton potential and a range $\lambda$ related through a Yukawa-like relation to
the mass scale of the hypothetical new particle which would mediate the ``fifth force''.
A recent update of the status of such a fifth force is shown on Fig.~1 in~\cite{Jaekel05},
reproduced thanks to a courtesy of J.~Coy, E.~Fischbach, R.~Hellings, C.~Talmadge and
E.M.~Standish. It shows that the Yukawa term is excluded with a high accuracy
at ranges tested in lunar laser ranging~\cite{Williams96}
and tracking of martian probes~\cite{Anderson96}.
At the same time, it also makes it clear that windows remain open for large corrections
($\alpha>1$) at short ranges as well as long ranges.

The short range window is being actively explored, with laboratory experiments
reaching an impressive accuracy at smaller and smaller distances~\cite{Kapner07}.
At even shorter ranges, tests of the gravity force law are pursued as careful
comparaisons between theoretical predictions and experimental measurements
of the Casimir force, which becomes dominant at micrometric
ranges~\cite{Onofrio06,Lambrecht06}.
In the long range window, a test of the gravity force was initiated by NASA
as the extension of Pioneer 10/11 missions after their primary planetary
objectives had been met.
This led to the largest scaled test of gravity ever carried out,
with the striking output of a signal that failed to
confirm the known laws of gravity~\cite{Anderson98,Anderson02}.

This so-called ``Pioneer anomaly'' was recorded on Doppler tracking data
of the Pioneer 10 \& 11 probes by the NASA deep space network.
It is thus a result of radionavigation techniques,
which are based on the performances of the accurate reference clocks
located at reception and emission stations~\cite{Asmar05}.
The Doppler observable can equivalently be interpreted as
a relative velocity of the probe with respect to the station, which contains
not only the effect of motion but also relativistic and gravitational effects.
The anomaly has been registered on the
two deep space probes showing the best navigation accuracy.
This is not an impressive statistics when compared to the large number
of tests confirming GR.
In particular, when the possibility of an artefact onboard the probe
is considered, this artefact could be the same on the two probes.
A number of mechanisms have been considered as attempts of explanations of the
anomaly as a systematic effect generated  by the spacecraft itself or its
environment (see the references in~\cite{Nieto05}) but they have not
led to a satisfactory understanding to date.

The Pioneer anomaly constitutes an intriguing piece of information in a
context where the status of gravity theory
is challenged by the puzzles of dark matter and dark energy.
If confirmed, this signal might reveal an anomalous behaviour of gravity
at scales of the order of the size of the solar system and thus have a
strong impact on fundamental physics, solar system physics, astrophysics
and cosmology.
It is therefore important to use as many investigation techniques as
might be available for gaining new information.

The Pioneer data which have shown an anomaly have been re-analyzed
by several independent groups which have confirmed the presence
of the anomaly~\cite{Markwardt02,Olsen07,Levy09}.
As data covering the whole period of Pioneer 10 \& 11 missions from
launch to the last data point have been saved, it is worth
investigating not only the Doppler tracking data, but also
the telemetry data~\cite{Turyshev06}.
These efforts should lead to an improved control of systematics
and produce new information of importance on several properties of the force,
for example its direction, long-term variation as well as annual or diurnal
modulations, spin dependence, a question of particular interest being
that of the onset of the anomaly.

If there exist gravity theories where a Pioneer signal can take a natural place,
they must be considered with great care.
If on the contrary one can prove that there exist no such
theories, this result is also important since it allows the
range of validity of GR to be extended to the size of the solar system.
Anyway, the relevance of the Pioneer anomaly for space navigation is already
sufficient to deserve a close scrutiny.
This means that the outputs of the data analysis have to be compared
with the predictions of possible theoretical explanations of
the Pioneer anomaly
(see for example~\cite{Jaekel06,Brownstein06,Bruneton07,Reynaud09}
and references therein).

In the meantime, new missions have been designed to study the anomaly and understand
its origin. In particular, two missions have been proposed to the Cosmic Vision process
at ESA, the first one as a medium size mission~\cite{Christophe09} using accelerometer and
radioscience instruments upgraded from existing technology,
the second one as a large size mission using new quantum sensors to map the
two components of the gravity field with high accuracy~\cite{Wolf09}.
These projects are based on new technologies
of importance for future fundamental physics and deep space exploration.
In the following, we will focus our attention on atomic clocks,
as used in the projects ACES~\cite{Salomon01} and SAGAS~\cite{Wolf09}.
We will also discuss their applications,
not only for tests in fundamental physics,
but also for other important purposes such as
earth sciences, solar system physics or navigation.

\section{Atomic Clock Ensemble in Space (ACES)}

The measurement of time intervals has experienced spectacular progress over the last centuries. In the middle of the
$20^{th}$ century, the invention of the quartz oscillator and the first atomic clocks opened a new era for time
keeping, with inacuracies improving from 10 microseconds per day in 1967 to 100 picoseconds per day for the primary
cesium atomic clocks using laser cooled atomic fountains. The best atomic fountains approach 10 picosecond error per
day, i.e a frequency stability of 1 part in $10^{16}$ \cite{Bize05} while the most recent atomic clocks, operating in
the optical domain, reach 2 picoseconds per day \cite{Rosenband08,Ludlow08} and improve at a fast pace.

Because time intervals and frequencies can be measured so precisely, applications of atomic clocks are numerous and
diverse. Most precision measurements and units of the SI system can be traced back to frequencies. With the
redefinition of the meter 25 years ago and the choice of a conventional value for the speed of light in vacuum,
distance measurements have been simply translated into time interval measurements. In other words, the development of
quantum technologies has led to large progress in the investigation of our spatio-temporal environment. This is
illustrated by the impressive improvement of the measurement of the Einstein redshift effect \cite{Vessot80,Pound99},
which will be pushed further by the ACES project  (see the discussion below). This is also made clear by the
measurements of distances in the solar system, with the astronomical unit now connected to atomic units and
consequently known with a much better accuracy than previously \cite{Shapiro99}.

The most visible consequence of these ideas is given by the Global Navigation Satellite Systems (GNSS), which are a
spectacular and initially unexpected application of the resources of quantum physics (the precision of the system comes
from the atomic clocks in the satellites) and general relativity \cite{Ashby03}. The US Global Positioning System (GPS)
enables today any user with a small receiver or modern cell phone to locate its position on the globe with meter
accuracy. In a few years from now, the GALILEO system will be the european contribution to GNSS and its vast uses for
various applications, in particular in the scientific domain in geodesy, Earth monitoring, and time metrology.

In the following, we review some of the properties of space clocks in the context of the ACES mission (Atomic Clock
Ensemble in Space). ACES is an ESA mission in fundamental physics \cite{Salomon01,Cacciapuoti07,Salomon07}. ACES aims
at flying a new generation of atomic clocks onboard the International Space Station (ISS) and comparing them to a
network of ultra-stable clocks on the ground. The ISS is orbiting at a mean elevation of 400 km with 90 min. of
rotation period and an inclination angle of $51.6^\circ$. ACES will be transported on orbit by the Japanese transfer
vehicle HTV, and installed at the external payload facility of the Columbus module using the ISS robotic arm.

The ACES payload accommodates two atomic clocks: PHARAO, a primary frequency standard based on samples of laser cooled
cesium atoms, and the active hydrogen maser SHM. The performances of the two clocks are combined to generate an
on-board timescale with the short-term stability of SHM and the long-term stability and accuracy of the cesium clock
PHARAO. The on-board comparison of PHARAO and SHM and the distribution of the ACES clock signal are ensured by the
Frequency Comparison and Distribution Package (FCDP), while all data handling processes are controlled by the eXternal
PayLoad Computer (XPLC). A GNSS receiver installed on the ACES payload and connected to the on-board time scale will
provide precise orbit determination of the ACES clocks.  The ACES clock signal will be transferred on ground by a time
and frequency transfer link in the microwave domain (MWL). MWL compares the ACES frequency reference to a set of ground
clocks, enabling fundamental physics tests and applications in different areas of research.

The planned mission duration is 18 months with a possible extension to 3 years. During the first two months, the
functionality of the clocks and of MWL will be tested. Then, a period of 4 months will be devoted to the performance
evaluation of the clocks. During this phase, a signal with frequency inaccuracy in the $10^{-15}$ range will be
available to ground users. In microgravity, the linewidth of the atomic resonance of the PHARAO clock will be tuned by
two orders of magnitude, down to sub-Hertz values (from 11 Hz to 110 mHz), 5 times narrower than in Earth based atomic
fountains. After clock optimization, performances in the $10^{-16}$ range are expected both for  frequency instability
and inaccuracy. In the second part of the mission (12 to 30 months), the on-board clocks will be compared to a network
of ground atomic clocks operating both in the microwave and optical domain.

The scientific objectives of the ACES mission  cover a wide spectrum. The frequency comparisons between the space
clocks and ground clocks will be used to test general relativity to high accuracy. As recalled above identical clocks
located  at different positions in gravitational fields experience a frequency shift that depends directly on the
component $g_{00}$ of the metric, i.e to the Newtonian potential at the clock position. The comparison between the ACES
onboard clocks and ground-based atomic clocks will measure the frequency variation due to the gravitational red-shift
with a 70-fold improvement on the best previous experiment \cite{Vessot80}, testing the Einstein prediction at the 2
ppm uncertainty level. Time variations of fundamental constants will be measured by comparing clocks based on different
transitions or different atomic species \cite{Fortier07}. Any transition energy in an atom or a molecule can be
expressed in terms of the fine structure constant $\alpha$ and the two dimensionless constants $m_q/\Lambda_{QCD}$ and
$m_e/\Lambda_{QCD}$, involving the quark mass $m_q$, the electron mass $m_e$ and the QCD mass scale $\Lambda_{QCD}$
\cite{Flambaum04,Flambaum06}. ACES will perform crossed comparisons of ground clocks both in the microwave and in the
optical domain with a resolution of $10^{-17}$ after a few days of integration time. These comparisons will impose
strong and unambiguous constraints on time variations of fundamental constants reaching an uncertainty of
$10^{-17}$/year in case of a 1-year mission duration and $3\,10^{-18}$/year after three years. ACES will also perform
tests of the Local Lorentz Invariance (LLI), according to which the outcome of any local test experiment is independent
of the velocity of the freely falling apparatus. In 1997, LLI tests based on the measurement of the round-trip speed of
light have been performed by comparing clocks on-board GPS satellites to ground hydrogen masers \cite{Wolf97}. In such
experiments, LLI violations would appear as variations of the speed of light $c$ with the direction and the relative
velocity of the clocks. ACES will perform a similar experiment by measuring relative variations of the light velocity
at the $10^{-10}$ uncertainty level.

Developed by CNES, the cold atom clock PHARAO will combine laser cooling techniques and microgravity conditions to
significantly increase the interaction time and consequently reduce the linewidth of the clock transition. Improved
stability and better control of systematic effects will be demonstrated in the space environment. PHARAO will reach a
fractional frequency instability of $1\cdot10^{-13}\cdot\tau^{-1/2}$, where $\tau$ is the integration time expressed in
seconds, and an inaccuracy of a few parts in $10^{16}$. The engineering model of the PHARAO clock has been completed
and is presently under test at CNES premises in Toulouse. Design and first results are presented in \cite{Laurent06}.

Developed by SPECTRATIME under ESA coordination, SHM provides ACES with a stable
fly-wheel oscillator. The main challenge of SHM is represented by the low mass
and volume figures required by the space clock with respect to ground
H-masers. SHM will provide a clock signal with fractional
frequency instability down to $1.5\cdot10^{-15}$ after only $10^4$~s of
integration time. Two servo-loops will lock together the clock signals of PHARAO
and SHM generating an on-board time scale combining the short-term stability of
the H-maser with the long-term stability and accuracy of the cesium clock.
The Frequency Comparison and Distribution Package (FCDP) is the central node of
the ACES payload. Developed by ASTRIUM and TIMETECH under ESA coordination,
FCDP is the on-board hardware which compares the signals delivered by the two
space clocks, measures and optimizes the performances of the ACES frequency
reference, and finally distributes it to the ACES microwave link MWL.

With the microwave link MWL, frequency transfer with time
deviation better than 0.3~ps at 300~s, 7~ps at 1~day, and 23 ps at 10 days
of integration time will be demonstrated.
The relativistic treatment of space to ground time transfer must be done up to
third order in $v/c$ in order to achieve sub $10^{-16}$ accuracy in clock
comparisons \cite{Blanchet01}. The gravitational shift measurement also requires
precise orbit determination and knowledge of the Earth gravitational potential
as described in \cite{Duchayne07}. MWL is developed by ASTRIUM, KAYSER-THREDE
and TIMETECH under ESA coordination. The proposed MWL concept is an upgraded
version of the Vessot two-way technique used for the GP-A experiment \cite{Vessot80}
and the PRARE geodesy instrument.

The frequency resolution that the ACES microwave link should reach in operational conditions is surpassing by one to
two orders of magnitude the existing satellite time transfer comparison methods based on the GPS system and TWSTFT (Two
Way Satellite Time and Frequency Transfer) \cite{Bauch06}. It may be compared with T2L2 (Time Transfer by Laser Link)
\cite{Exertier08} which is now flying onboard the JASON-2 satellite and taking science data since June 2008. The
assessment of its time transfer capability is presently ongoing with the interesting prospect to reach a time
resolution of $\simeq 10\,$ps after one day of averaging for clock comparisons in common view. For the ACES mission,
due to the low orbit of the ISS and the limited duration of each pass (300-400\,s) over a given ground station, the
common view technique will be suitable for comparing ground clocks over continental distances, for instance within
Europe or USA or Japan. In a common view comparison, the frequency noise of the onboard clock cancels out to a large
degree so that only the link instability remains. ACES mission will also demonstrate the capability to compare ground
clocks in non common view with a resolution better than $10^{-13} \Delta t^{1/2}$ for $\Delta t > 1000$\,s, that is 3
ps and 10 ps for space-ground comparisons separated by 1000 s and 10000 s respectively. This science objective takes
full benefit of the excellent onboard time scale realized by the combination of SHM and PHARAO.

These performances will enable common view and non-common view comparisons of ground clocks with $10^{-17}$ frequency
resolution after few days of integration time. The recent development of optical frequency combs
\cite{Holzwarth00,Diddams00} awarded with the 2005 Nobel Prize in Physics, significantly simplifies the link between
optical and microwave frequencies. From this point of view, ACES will take full advantage of the recent progress of
optical clocks~\cite{Rosenband08,Ludlow08}, reaching today instability and inaccuracy levels of 2 parts in $10^{17}$.
Multiple frequency comparisons between a variety of advanced ground clocks will be possible among the 35 institutes
which have manifested their interest to participate to the ACES mission. This is important for the tests of the
variability of fundamental physical constants, and international comparisons of time scales.

Other applications of the ACES clock signal are currently being developed. ACES will demonstrate a new ``relativistic
geodesy'' based on a differential measurement of the Einstein's gravitational red-shift between distant ground clocks.
It will take advantage of the accuracy of ground-based optical clocks to resolve differences in the Earth gravitational
potential at the $\simeq 10$~cm level. A 10 cm change in elevation on the ground amounts to a frequency shift of 1 part
in $10^{17}$, compared to today's optical clock accuracy of 2 parts in $10^{17}$. The GNSS receiver on-board the ACES
payload will allow to monitor the GNSS networks and develop interesting applications in Earth remote sensing. This
includes studies of the oceans surface via GNSS reflectometry measurements and the analysis of the Earth atmosphere
with GNSS radio-occultation experiments.

In addition, ACES will deliver a
global atomic time scale with $10^{-16}$ accuracy, it will allow clocks synchronization
at an uncertainty level of 100~ps, and contribute to international atomic time scales.
At the $10^{-18}$ level, ground clocks will be limited by the fluctuations of the Earth potential
suggesting that future high precision time references will have to be placed in the
space environment where these fluctuations are significantly reduced. From this point of view,
ACES is the pioneer of new concepts for global time keeping and positioning based on a reduced
set of ultra-stable space clocks in orbit.

\section{SAGAS}

The SAGAS mission will study all aspects of large scale gravitational phenomena in the Solar System
using quantum technology, with science objectives in fundamental physics and Solar System exploration.
The large spectrum of science objectives makes SAGAS a unique combination of exploration and science,
with a strong basis in both programs. The involved large distances (up to 53 AU) and corresponding
large variations of gravitational potential combined with the high sensitivity of SAGAS instruments
serve both purposes equally well. For this reason, SAGAS brings together traditionally distant
scientific communities ranging from atomic physics through experimental gravitation to planetology
and Solar System science.

The payload will include an optical atomic clock optimised for long term performance, an absolute
accelerometer based on atom interferometry and a laser link for ranging, frequency comparison
and communication. The complementary instruments will allow highly sensitive measurements of all
aspects of gravitation via the different effects of gravity on clocks, light, and the free fall
of test bodies, thus effectively providing a detailed gravitational map of the outer Solar System
whilst testing all aspects of gravitation theory to unprecedented levels.

The SAGAS accelerometer is based on cold Cs atom technology derived to a large extent from the
PHARAO space clock built for the ACES mission discussed in the preceding section.
The PHARAO engineering model has recently been tested with success, demonstrating the expected
performance and robustness of the technology. The accelerometer will only require parts of PHARAO
(cooling and trapping region) thereby significantly reducing mass and power requirements.
The expected sensitivity of the accelerometer is $1.3\cdot10^{-9}$m/s$^2/\sqrt{{\Hz}}$
with an absolute accuracy (bias determination) of $5\cdot10^{-12}$m/s$^2$.

The SAGAS clock will be an optical clock based on trapped and laser cooled single ion technology
as pioneered in numerous laboratories around the world. In the present proposal it will be based
on a Sr$^{+}$ ion with a clock wavelength of 674 nm. The expected stability of the SAGAS clock is
$1\cdot10^{-14}/\sqrt{\tau}$ (with $\tau$ the integration time), with an accuracy in realising
the unperturbed ion frequency of $1\cdot10^{-17}$. The best optical single ion ground clocks
presently show stabilities below $4\cdot10^{-15}/\sqrt{\tau}$, slightly better than the one
assumed for the SAGAS clock, and only slightly worse accuracies $2\cdot10^{-17}$.
So the technology challenges facing SAGAS are not so much the required performance, but the
development of reliable and space qualified systems, with reduced mass and power consumption.

The optical link is using a high power (1 W) laser locked to the narrow and stable frequency provided by the optical
clock, with coherent heterodyne detection on the ground and on board the spacecraft. It serves the multiple purposes of
comparing the SAGAS clock to ground clocks, providing highly sensitive Doppler measurements for navigation and science,
and allowing data transmission together with timing and coarse ranging. It is based on a 40 cm space telescope and 1.5
m ground telescopes (similar to lunar laser ranging stations). The short term performance of the link in terms of
frequency comparison of distant clocks will be limited by atmospheric turbulence at a few $10^{-13}/\tau$, thus
reaching the clock performance after about 1000\,s integration time, which is amply sufficient for the SAGAS science
objectives. The main challenges of the link will be the required pointing accuracy (0.3$^{\prime\prime}$) and the
availability of space qualified, robust 1 W laser sources at 674 nm. Quite generally, laser availability and
reliability will be the key to achieving the required technological performances, for the clock as well as the optical
link.

For this reason a number of different options have been considered for the clock/link laser wavelength,
with several other ions that could be equally good candidates (e.g. Yb$^{+}$ at 435 nm and
Ca$^{+}$ at 729 nm). Given present laser technology, Sr$^{+}$ was preferred, but this choice could be
revised depending on laser developments over the next years. We also acknowledge the possibility
that femtosecond laser combs might be developed for space applications in the near future, which
would open up the option of using either ion with existing space qualified 1064 nm Nd:YAG lasers for the link.

More generally, SAGAS technology takes advantage of the important heritage from cold atom technology used in
PHARAO and laser link technology designed for LISA (Laser Interferometric Space Antenna) \cite{Danzmann03}.
It will provide an excellent opportunity to develop those technologies for general use, including development
of the ground segment (Deep Space Network telescopes and optical clocks), that will allow such technologies
to be used in many other mission configurations for precise timing, navigation and broadband data transfer
throughout the Solar System.

SAGAS will carry out a large number of tests of fundamental physics, and gravitation in particular, at scales
only attainable in a deep space experiment. The unique combination of onboard instruments will allow 2 to 5
orders of magnitude improvement on many tests of special and general relativity, as well as a detailed
exploration of a possible anomalous scale dependence of gravitation. It will also provide detailed information
on the Kuiper belt mass distribution and determine the mass of Kuiper belt objects and possibly discover new ones.
During the transits, the mass and mass distribution of the Jupiter system will be measured with unprecedented
accuracy. The science objectives are discussed in the following,
based on estimated measurement uncertainties of the different observables.

SAGAS will provide three fundamental measurements: the accelerometer readout and the two frequency differences
(measured on ground and on board the satellite) between the incoming laser signal and the local optical clock.
Auxiliary measurements are the timing of emitted/received signals on board and on the ground, which are used
for ranging and time tagging of data. The high precision science observables will be deduced from the fundamental
measurements by combining the measurements to obtain information on either the frequency difference between the
clocks or the Doppler shift of the transmitted signals. The latter gives access to the relative satellite-ground
velocity, from which the gravitational trajectory of the satellite can be deduced by correcting non-gravitational
accelerations using the accelerometer readings.

In the following, we assume that Earth station motion and its local gravitational potential can be known and
corrected to uncertainty levels below 10$^{-17}$ in relative frequency (10 cm on geocentric distance),
which, although challenging, are within present capabilities. For the Solar System parameters this requires
10$^{-9}$ relative uncertainty for the ground clock parameters ($GM$ and $r$ of Earth), also achieved at
present \cite{Groten99}, and less stringent requirements for the satellite.

For long term integration and the determination of an acceleration bias, the limiting factor will then be the
accelerometer noise and absolute uncertainty (bias determination). More generally, modelling of non-gravitational
accelerations will certainly allow some improvement on the long term limits imposed by the accelerometer noise
and absolute uncertainty, but is not taken into account in our preliminary evaluation \cite{Wolf09}.
Also, depending on the science goal and corresponding signal, the ground and on-board data can be combined in a
way to optimise the S/N ratio (see \cite{Reynaud08} for details). For example, in the Doppler observable, the
contribution of one of the clocks (ground or space) can be made negligible by combining signals such that one
has coincidence of the ``up'' and ``down'' signals on board or on the ground.

We will use a mission profile with a nominal mission lifetime of 15 years and the possibility of an extended
mission to 20 years if instrument performance and operation allow this. In that time frame, the trajectory
allows the satellite to reach a heliocentric distance of 39 AU in nominal mission and 53 AU with extended duration.

In General Relativity (GR), the frequency difference of two ideal clocks is proportional to
(see eq.(\ref{Deltas}) to first order in the weak field approximation, with $g_{00}\simeq 1-2w/c^2$
and $g_{rr}\simeq -1$; $w$ is the Newtonian gravitational potential
and $v$ the coordinate velocity)
\begin{eqnarray}
\label{clock_signal}
&& \frac{\d s_\g}{c\d t} - \frac{\d s_\s}{c\d t} \simeq
\frac{w_\g-w_\s}{c^2} +  \frac{v^2_\g-v^2_\s}{2c^2}
\end{eqnarray}
In theories different from GR this relation is modified, leading to different time and space dependence of the
frequency difference. This can be tested by comparing two clocks at distant locations (different values of $w$ and $v$)
via exchange of an electromagnetic signal.  The SAGAS trajectory (large potential difference) and low uncertainty
on the observable (\ref{clock_signal}) allows a relative uncertainty on the redshift determination given by the
10$^{-17}$ clock bias divided by the maximum value of $\Delta w/c^2$. For a distance of 50 AU this corresponds
to a test with a relative uncertainty of $1\cdot10^{-9}$, an improvement by almost 5 orders of magnitude on the
uncertainty obtained by the most sensitive experiment at present \cite{Vessot80}.

Additionally, the mission also provides the possibility of testing the velocity term in (\ref{clock_signal}),
which amounts to a test of Special Relativity (Ives-Stilwell test), and thus of Lorentz invariance.
Towards the end of the nominal mission, this term is about $4\cdot10^{-9}$, and can therefore be measured by SAGAS
with $3\cdot10^{-9}$ relative uncertainty. The best present limit on this type of test is $2\cdot10^{-7}$
\cite{Saathoff03}, so SAGAS will allow an improvement by a factor $\sim70$. Considering a particular preferred
frame, usually taken as the frame in which the 3K cosmic background radiation is isotropic, one can set an even
more stringent limit. In that case a putative effect will be proportional to $(\bv_\s-\bv_\g).\bv_\S/c^2$
(see \cite{Saathoff03}), where $\bv_\s$ and $\bv_\g$ are the velocity vectors of the satellite and ground
while $\bv_\S$ is the velocity of the Sun through the CMB frame ($\sim350$ km/s). Then SAGAS will allow a
measurement with about $5\cdot10^{-11}$ relative uncertainty, which corresponds to more than 3 orders of
magnitude improvement on the present limit. Note that Ives-Stilwell experiments also provide the best present
limit on a particularly elusive parameter ($\varkappa_\mathrm{tr}$) of the Lorentz violating Standard Model
Extension (SME) photon sector \cite{Hohensee07}, so that SAGAS also allows for the same factor 70 to 1000
improvement on that parameter.

Spatial and/or temporal variations of fundamental constants constitute another violation of Local Position
Invariance and thus of GR. Over the past few years, there has been great interest in that possibility
(see e.g. \cite{Uzan03} for a review), spurred on the one hand by models for unification theories of the
fundamental interactions where such variations appear quite naturally, and on the other hand by recent
observational claims of a variation of different constants over cosmological timescales \cite{Murphy03,Reinhold06}.
Such variations can be searched for with atomic clocks, as the involved transition frequencies depend on
combinations of fundamental constants and in particular, for the optical transition of the SAGAS clock,
on the fine structure constant $\alpha$. Such tests take two forms: searches for a drift in time of
fundamental constants, or for a variation of fundamental constants with ambient gravitational field.
The latter tests for a non-universal coupling between ambient gravity and non-gravitational interactions
(clearly excluded by GR) and is well measured by SAGAS, because of the large change in gravitational
potential during the mission.

For example, some well-known string theory based models associates a scalar field such as the dilaton to
the standard model \cite{Damour93,Damour94,Flambaum07}. Such scalar fields would couple to ordinary matter
and thus their non-zero value would introduce a variation of fundamental constants, in particular $\alpha$
of interest here. The non-zero value of such scalar fields could be of cosmological origin, leading to a
constant drift in time of fundamental constants, and/or of local origin, i.e. taking ordinary matter as
its source \cite{Flambaum07}. In the latter case one would observe a variation of fundamental constants
with the change in local gravitational potential $w$, which can be simply written as
$\delta\alpha/\alpha=k_\alpha\delta w/c^2$.
The difference in gravitational potential between the Earth and the SAGAS satellite at the end of nominal mission
is about $\delta w/c^2\simeq10^{-8}$, which is 30 times more than the variation attainable on Earth.
With a Sr$^{+}$ optical transition used in the SAGAS clock and a ground clock with 10$^{-17}$ uncertainty,
this yields a limit of $k_\alpha< 2.4\cdot10^{-9}$, a factor 250 improvement over the best present
limit \cite{Flambaum07}.

SAGAS also offers a possibility to improve the test of the PPN family of metric theories of gravitation
\cite{Will01}. The two most common parameters of the PPN framework are the Eddington parameters $\beta$
and $\gamma$. Present limits on $\gamma$ are obtained from measurements on light
propagation such as light deflection, Shapiro delay and Doppler velocimetry
of the Cassini probe during the 2002 solar occultation \cite{Bertotti03}.
SAGAS will carry out similar measurements during solar conjunctions, with improved sensitivity and at
optical rather than radio frequencies, which significantly minimizes errors due to the solar corona and
the Earth's ionosphere.
When combining the on board and ground measurements such that the ``up'' and ``down'' signals coincide
at the satellite (classical Doppler type measurement), the noise from the on-board clock cancels
to a large extent and one is left with noise from the accelerometer, the ground clock, and the atmosphere.
Details on these noise sources, and on the effects of atmospheric turbulence, variations in temperature,
pressure and humidity are presented in \cite{Wolf09}. The resulting improvement on the estimation
of $\gamma$ is of the order of 100, with some potential for further improvement if several
occultations may be analysed.

As already discussed, experimental tests of gravity have shown a good overall agreement with GR,
but most theoretical models aimed at inserting GR within the quantum framework predict observable
modifications at large scales. The anomalies observed on the Pioneer probes, as well as
the phenomena commonly ascribed to ``dark matter'' and ``dark energy'', suggest that it is
extremely important to test the laws of gravity at interplanetary distances \cite{Reynaud09}.
This situation has motivated an interest in flying new probes to the distances where the anomaly
was first discovered, that is beyond the Saturn orbit, and studying gravity with the largely
improved spatial techniques which are available nowadays.

SAGAS has the capability to improve our knowledge of the law of gravity at the scale of the
solar system, and to confirm or infirm the presence of a Pioneer anomaly (PA).
With one year of integration, all SAGAS observables allow a measurement of any effect of the size of
the PA with a relative uncertainty of better than 1\%. This will allow a "mapping" of any
anomalous scale dependence over the mission duration and corresponding distances.
Furthermore, the complementary observables available on SAGAS allow a good discrimination between
different hypotheses thereby not only measuring a putative effect but also allowing an identification
of its origin. SAGAS thus offers the possibility to constrain a significant number of theoretical
approaches to scale dependent modifications of GR. Given the complementary observables available on
SAGAS the obtained measurements will provide a rich testing ground for such theories with the potential
for major discoveries.

Let us emphasize at this point that this gravity law test is important not only for fundamental physics
but also for solar system science. The Newton law of gravity is indeed a key ingredient of the models
devoted to a better understanding of the origin of the solar system.
The Kuiper belt (KB) is a collection of masses, remnant of the circumsolar disk where giant planets of
the solar system formed 4.6 billion years ago. Precise measurements of its mass distribution would
significantly improve our understanding of planet formation not only in the solar system but also
in recently discovered planetary systems. The exceptional sensitivity and versatility of SAGAS
for measuring gravity can be used to study the sources of gravitational fields in the outer solar system,
and in particular the class of Trans Neptunian Objects (TNOs), of which those situated in the KB have been
the subject of intense interest and study over the last years \cite{Morbidelli07}. Observation of KB objects
(KBOs) from the Earth is difficult due to their relatively small size and large distance, and estimates of
their masses and distribution are accordingly inaccurate. Estimates of the total KB mass from the
discovered objects ($\sim1000$ KBOs) range from 0.01 to 0.1 Earth masses, whereas in-situ formation of
the observed KBOs would require three orders of magnitude more solid material in a dynamically cold disk.

A dedicated probe like SAGAS will help discriminating different models of the spatial distribution of the KB
and for determining its total mass. The relative frequency shift between the ground and space clock due to the KB
gravitational potential is indeed a sensitive probe of the spatial distribution of this mass
(see \cite{Bertolami06,Bertolami07,Wolf09} for details).
The SAGAS frequency observable is well suited to study the large, diffuse, statistical mass distribution of
KBOs essentially due to its sensitivity directly to the gravitational potential (1/r dependence), rather than
the acceleration (1/r$^2$ dependence). The large diffuse signal masks any signal from individual KBOs.
When closely approaching one of the objects, the crossover between acceleration sensitivity
(given by the $5\cdot10^{-12}$ m/s$^2$ uncertainty on non gravitational acceleration) and the frequency
sensitivity ($10^{-17}$ uncertainty on $w/c^2$) for an individual object is situated at about 1.2 AU.
Below that distance the acceleration measurement is more sensitive than the frequency measurement.
This suggests a procedure to study individual objects using the SAGAS observables: use the satellite
trajectory (corrected for the non gravitational acceleration) to study the gravity from a close object
and subtract the diffuse background from all other KBOs using the frequency measurement. Investigating
several known KBOs within the reach of SAGAS \cite{Bernstein04} shows that their masses can be determined
at the \% level when approaching any of them to 0.2 AU or less. Of course, this also opens the way towards
the discovery of new such objects, too small to be visible from the Earth.
Similarly, during a planetary flyby the trajectory determination (corrected for the non gravitational
acceleration) will allow the determination of the gravitational potential of the planetary system.
The planned Jupiter flyby with a closest approach of $\sim600000$ km will improve present knowledge of
Jovian gravity by more than two orders of magnitude.

Doppler ranging to deep space missions provides the best upper limits available at present on
gravitational waves (GW) with frequencies of order $c/L$ where $L$ is the spacecraft to ground distance
i.e. in the 0.01 to 1 mHz range \cite{Armstrong87,Armstrong03}, and even down to 1 $\mu$Hz,
albeit with lower sensitivity \cite{Anderson85,Armstrong03}. The corresponding limits on GW are
determined by the noise PSD of the Doppler ranging to the spacecraft for stochastic GW backgrounds,
filtered by the bandwidth of the observations when looking for GW with known signatures.
In the case of SAGAS data, this yields a strain sensitivity of $10^{-14}/\sqrt{{\Hz}}$ for
stochastic sources in the frequency range of 0.06 to 1 mHz with a $f^{-1}$ increase at low
frequency due to the accelerometer noise. When searching for GW with particular signatures in this
frequency region, optimal filtering using a corresponding GW template will allow reaching strain
sensitivities as low as $10^{-18}$ with one year of data. This will improve on best present upper
limits on GW in this frequency range by about four orders of magnitude.
Even if GW with sufficiently large amplitudes are not found, still the results might serve as
upper bounds for astrophysical models of known GW sources \cite{Reynaud08}.


{
\acknowledgements
\footnotesize
We are indebted to a large number of colleagues with whom we have
discussed the topics of this paper over the years. 
Special thanks are due to CNES for its long term support.
}

\def\sci{{Science}}
\def\ag{{Ann.\ Geo\-phy\-si\-c\ae}}
\def\asr{{Adv.\ Space Res.}}
\def\apj{{Ast\-ro\-phys.\ J.}}
\def\apjl{{Ast\-ro\-phys.\ J.\ Lett.}}
\def\esa{{ESA--SP}}
\def\grl{{Geo\-phys.\ Res.\ Lett.}}
\def\ieg{{IEEE Trans.\ Geo\-sci.\ Re\-mo\-te Sens.}}
\def\iep{{IEEE Trans.\ Plasma Sci.}}
\def\jatp{{J.\ Atmos.\ Ter\-res\-tr.\ Phys.}}
\def\jgg{{J.\ Geo\-magn.\ Geo\-elec.}}
\def\jgr{{J.\ Geo\-phys. Res.}}
\def\nat{{Nature}}
\def\prl{{Phys.\ Rev.\ Lett.}}
\def\pr{{Phys.\ Rev.}}
\def\pss{{Planet.\ Space Sci.}}
\def\rgsp{{Rev.\ Geo\-phys.\ Space Phys.}}
\def\ssr{{Space Sci.\,Rev.}}
\def\rmp{{Rev.\ Mod.\ Phys.}}
\def\pl{{Phys.\ Lett.}}
\def\pf{{Phys. Fluids}}
\def\npg{{Nonlin.\ Process.\ Geophys.}}
\def\njp{{New J.\ Phys.}}
\def\pop{{Phys.\ Plasmas}}
\def\jpp{{J.\ Plasma Phys.}}
\def\mnras{{Monthly Notic.\ Royal Astron.\ Soc.}}
\def\cqg{{Class. Quantum Grav.}}
\def\ijmp{{Int. J. Mod. Phys.}}
\def\aa{{Astron. \& Astrop.}}
\def\ea{{Exp.\ Astron.}}

\vspace{-0.3cm}
\def\ibid{\textit{ibidem }}
\def\url#1{{#1}}
\def\arxiv#1{{#1}}
\def\REVIEW#1#2#3#4{\textit{#1} \textbf{#2} {#4} ({#3})}
\def\Book#1#2#3{\textit{#1} ({#2}, {#3})}
\def\BOOK#1#2#3#4{\textit{#1} ({#2}, {#3}, {#4})}
\def\BOOKed#1#2#3#4#5{\textit{#1}, #2 ({#3}, {#4}, {#5})}
\def\Name#1#2{#2~#1}

}
\end{article}


{\scriptsize
\begin{thebibliography}{000}


\bibitem[\protect\citeauthoryear{This Issue}{2009}]{ThisIssue}
This issue.

\bibitem[\protect\citeauthoryear{Adelberger}{2003}]{Adelberger03}
\Name{Adelberger}{E.G.}, \Name{Heckel}{B.R.} and \Name{Nelson}{A.E.},
\REVIEW{Ann. Rev. Nucl. Part. Sci.}{53}{2003}{77}.

\bibitem[\protect\citeauthoryear{Anderson}{1985}]{Anderson85}
\Name{Anderson}{J.D.} and \Name{Mashoon}{B.},
\REVIEW{\apj}{290}{1985}{445}.

\bibitem[\protect\citeauthoryear{Anderson}{1996}]{Anderson96}
\Name{Anderson}{J.D.} \etal,
\REVIEW{\apj}{459}{1996}{365}.

\bibitem[\protect\citeauthoryear{Anderson}{1998}]{Anderson98}
\Name{Anderson}{J.D.} \etal,
\REVIEW{\prl}{81}{1998}{2858}.

\bibitem[\protect\citeauthoryear{Anderson}{2002}]{Anderson02}
\Name{Anderson}{J.D.} \etal,
\REVIEW{\pr}{D65}{2002}{082004}.

\bibitem[\protect\citeauthoryear{Armstrong}{1987}]{Armstrong87}
\Name{Armstrong}{J.W.} \etal,
\REVIEW{\apj}{318}{1987}{536}.

\bibitem[\protect\citeauthoryear{Armstrong}{2003}]{Armstrong03}
\Name{Armstrong}{J.W.} \etal,
\REVIEW{\apj}{599}{2003}{806}.

\bibitem[\protect\citeauthoryear{Ashby}{2003}]{Ashby03}
\Name{Ashby}{N.},
\REVIEW{Living Rev. Rel.}{}{2003}{1}
\url{http://www.livingreviews.org/lrr-2003-1}.

\bibitem[\protect\citeauthoryear{Ashby}{2007}]{Ashby07}
\Name{Ashby}{N.} \etal,
\REVIEW{\prl}{98}{2007}{070802}.

\bibitem[\protect\citeauthoryear{Asmar}{2005}]{Asmar05}
\Name{Asmar}{S.W.} \etal,
\REVIEW{Radio Science}{40}{2005}{RS2001}.

\bibitem[\protect\citeauthoryear{Bauch}{2006}]{Bauch06}
\Name{Bauch}{A.} \etal,
\REVIEW{Metrologia}{43}{2006}{109120}.

\bibitem[\protect\citeauthoryear{Bernstein}{2004}]{Bernstein04}
\Name{Bernstein}{G.} \etal,
\REVIEW{\apj}{128}{2004}{1364}.

\bibitem[\protect\citeauthoryear{Bertolami}{2006}]{Bertolami06}
\Name{Bertolami}{O.} and \Name{Vieira}{P.},
\REVIEW{\cqg}{23}{2006}{4625}.

\bibitem[\protect\citeauthoryear{Bertolami}{2007}]{Bertolami07}
\Name{Bertolami}{O.} and \Name{Páramos}{J.},
\REVIEW{\ijmp}{D16}{2007}{1611}.

\bibitem[\protect\citeauthoryear{Bertotti}{2003}]{Bertotti03}
\Name{Bertotti}{B.}, \Name{Iess}{L.} and \Name{Tortora}{P.},
\REVIEW{Nature}{425}{2003}{374}.

\bibitem[\protect\citeauthoryear{Bize}{2003}]{Bize03}
\Name{Bize}{S.} \etal,
\REVIEW{\prl}{90}{2003}{150802}.

\bibitem[\protect\citeauthoryear{Bize}{2005}]{Bize05}
\Name{Bize}{S.} \etal,
\REVIEW{J. Phys.}{B38}{2005}{S449}.

\bibitem[\protect\citeauthoryear{Blanchet}{2001}]{Blanchet01}
\Name{Blanchet}{L.} \etal,
\REVIEW{\aa}{370}{2001}{320}.

\bibitem[\protect\citeauthoryear{Brownstein}{2006}]{Brownstein06}
\Name{Brownstein}{J.R.} and \Name{Moffat}{J.W.},
\REVIEW{\cqg}{23}{2006}{3427}.

\bibitem[\protect\citeauthoryear{Bruneton}{2007}]{Bruneton07}
\Name{Bruneton}{J.P.} and \Name{Esposito-Farèse}{G.},
\REVIEW{\pr}{D76}{2007}{124012 \& 129902}.

\bibitem[\protect\citeauthoryear{Cacciapuoti}{2007}]{Cacciapuoti07}
\Name{Cacciapuoti}{L.} \etal,
\REVIEW{Nuclear Physics}{B166}{2007}{303}.

\bibitem[\protect\citeauthoryear{Christophe}{2009}]{Christophe09}
\Name{Christophe}{B.} \etal,
\REVIEW{\ea}{23}{2009}{529}.

\bibitem[\protect\citeauthoryear{Damour}{1993}]{Damour93}
\Name{Damour}{T.} and \Name{Nordtvedt}{K.},
\REVIEW{\prl}{70}{1993}{2217};
\REVIEW{\pr}{D48}{1993}{3436}.

\bibitem[\protect\citeauthoryear{Damour}{1994}]{Damour94}
\Name{Damour}{T.} and \Name{Polyakov}{A.M.},
\REVIEW{Nucl. Phys.}{B423}{1994}{532}.

\bibitem[\protect\citeauthoryear{Damour}{2002}]{Damour02}
\Name{Damour}{T.}, \Name{Piazza}{F.} and \Name{Veneziano}{G.},
\REVIEW{\pr}{D66}{2002}{046007}.

\bibitem[\protect\citeauthoryear{Danzmann}{2003}]{Danzmann03}
\Name{Danzmann}{K.} \etal,
\REVIEW{\asr}{32}{2003}{1233}.

\bibitem[\protect\citeauthoryear{Diddams}{2000}]{Diddams00}
\Name{Diddams}{S.A.} \etal,
\REVIEW{\prl}{84}{2000}{5102}.

\bibitem[\protect\citeauthoryear{Duchayne}{2008}]{Duchayne07}
\Name{Duchayne}{L.}, \Name{Mercier}{F.} and \Name{Wolf}{P.},
\arxiv{0708.2387} (2007).

\bibitem[\protect\citeauthoryear{Einstein}{1907}]{Einstein07}
\Name{Einstein}{A.},
\REVIEW{Jahrbuch der Radioaktivit\"at und Elektronik}{4}{1907}{411}.

\bibitem[\protect\citeauthoryear{Einstein}{1911}]{Einstein11}
\Name{Einstein}{A.},
\REVIEW{Annalen der Physik}{35}{1911}{898}.

\bibitem[\protect\citeauthoryear{Einstein}{1915}]{Einstein15}
\Name{Einstein}{A.},
\REVIEW{Sitz. der Preuss. Akad. der Wissenschaften zu Berlin}{}{1915}{844}.

\bibitem[\protect\citeauthoryear{Einstein}{1916}]{Einstein16}
\Name{Einstein}{A.},
\REVIEW{Annalen der Physik}{49}{1916}{769}.

\bibitem[\protect\citeauthoryear{Exertier}{2008}]{Exertier08}
\Name{Exertier}{P.} \etal, IDS workshop, Nice, November 08,
http://ids.cls.fr/documents/report/ids~workshop~2008/IDS08~s8~Exertier~T2L2.pdf

\bibitem[\protect\citeauthoryear{Fischbach}{1998}]{Fischbach98}
\Name{Fischbach}{E.} and \Name{Talmadge}{C.},
\BOOK{The Search for Non Newtonian Gravity}{Springer Verlag}{Berlin}{1998}.

\bibitem[\protect\citeauthoryear{Fisher}{2004}]{Fisher04}
\Name{Fischer}{M.} \etal,
\REVIEW{\prl}{92}{2004}{230802}.

\bibitem[\protect\citeauthoryear{Flambaum}{2004}]{Flambaum04}
\Name{Flambaum}{V.V.} \etal,
\REVIEW{\pr}{D69}{2004}{115006}.

\bibitem[\protect\citeauthoryear{Flambaum}{2006}]{Flambaum06}
\Name{Flambaum}{V.V.} and \Name{Tedesco}{A.F.},
\REVIEW{\pr}{C73}{2006}{055501}.

\bibitem[\protect\citeauthoryear{Flambaum}{2007}]{Flambaum07}
\Name{Flambaum}{V.V.} and \Name{Shuryak}{E.V.},
arxiv{physics/0701220} (2007).

\bibitem[\protect\citeauthoryear{Fortier}{2007}]{Fortier07}
\Name{Fortier}{T.M.} \etal,
\REVIEW{\prl}{98}{2007}{070801}.

\bibitem[\protect\citeauthoryear{Groten}{1999}]{Groten99}
\Name{Groten}{E.}, Report of the IAG. Special Commission SC3,
Fundamental Constants, XXII IAG General Assembly (1999).

\bibitem[\protect\citeauthoryear{Hilbert}{1915}]{Hilbert}
\Name{Hilbert}{D.},
\REVIEW{Nachr. von der Gesellshaft der Wissenshaften zu G\"ottingen}{}{1915}{395}.

\bibitem[\protect\citeauthoryear{Hohensee}{2007}]{Hohensee07}
\Name{Hohensee}{M.} \etal,
\REVIEW{\pr}{D75}{2007}{049902}.

\bibitem[\protect\citeauthoryear{Holzwarth}{2000}]{Holzwarth00}
\Name{Holzwarth}{R.} \etal,
\REVIEW{\prl}{85}{2000}{2264}.

\bibitem[\protect\citeauthoryear{Jaekel}{2005}]{Jaekel05}
\Name{Jaekel}{M.-T.} and \Name{Reynaud}{S.},
\REVIEW{\ijmp}{A20}{2005}{2294}.

\bibitem[\protect\citeauthoryear{Jaekel}{2006}]{Jaekel06}
\Name{Jaekel}{M.-T.} and \Name{Reynaud}{S.},
\REVIEW{\cqg}{23}{2006}{777} \& 7561.

\bibitem[\protect\citeauthoryear{Kapner}{2007}]{Kapner07}
\Name{Kapner}{D.J.} \etal,
\REVIEW{\prl}{98}{2007}{021101}.

\bibitem[\protect\citeauthoryear{Lambrecht}{2006}]{Lambrecht06}
\Name{Lambrecht}{A.}, \Name{Maia Neto}{P.A.} and \Name{Reynaud}{S.},
\REVIEW{\njp}{8}{2006}{243}.

\bibitem[\protect\citeauthoryear{Laurent}{2006}]{Laurent06}
\Name{Laurent}{P.} \etal,
\REVIEW{Appl. Phys.}{B84}{2006}{683}.

\bibitem[\protect\citeauthoryear{Levy}{2009}]{Levy09}
\Name{Levy}{A.} \etal,
\REVIEW{\asr}{to be published}{2009}{}.

\bibitem[\protect\citeauthoryear{Ludlow}{2008}]{Ludlow08}
\Name{Ludlow}{A.D.} \etal,
\REVIEW{Science}{319}{2008}{1805}.

\bibitem[\protect\citeauthoryear{Marion}{2003}]{Marion03}
\Name{Marion}{H.} \etal,
\REVIEW{\prl}{90}{2003}{150801}.

\bibitem[\protect\citeauthoryear{Markwardt}{2002}]{Markwardt02}
\Name{Markwardt}{C.}, \arxiv{gr-qc/0208046};
\url{http://lheawww.gsfc.nasa.gov/users/craigm/atdf/}.

\bibitem[\protect\citeauthoryear{Mester}{2001}]{Mester01}
\Name{Mester}{J.} \etal,
\REVIEW{\cqg}{18}{2001}{2475}.

\bibitem[\protect\citeauthoryear{Morbidelli}{2007}]{Morbidelli07}
\Name{Morbidelli}{A.} in
\BOOKed{The Kuiper Belt}{A. Barruci \etal eds.}{Univ. of Arizona Press}{}{2007}.

\bibitem[\protect\citeauthoryear{Murphy}{2003}]{Murphy03}
\Name{Murphy}{M.T.} \etal,
\REVIEW{\mnras}{345}{2003}{609}.

\bibitem[\protect\citeauthoryear{Nieto}{2005}]{Nieto05}
\Name{Nieto}{M.M.}, \Name{Turyshev}{S.G.} and \Name{Anderson}{J.D.},
\REVIEW{\pl}{B613}{2005}{11}.

\bibitem[\protect\citeauthoryear{Olsen}{2007}]{Olsen07}
\Name{Olsen}{O.},
\REVIEW{\aa}{463}{2007}{393}.

\bibitem[\protect\citeauthoryear{Onofrio}{2006}]{Onofrio06}
\Name{Onofrio}{R.},
\REVIEW{\njp}{8}{2006}{237}.

\bibitem[\protect\citeauthoryear{Peik}{2004}]{Peik04}
\Name{Peik}{E.} \etal,
\REVIEW{\prl}{93}{2004}{170801}.

\bibitem[\protect\citeauthoryear{Petit}{2005}]{Petit05}
\Name{Petit}{G.} and \Name{Wolf}{P.},
\REVIEW{Metrologia}{42}{2005}{S138}.

\bibitem[\protect\citeauthoryear{Pound}{1999}]{Pound99}
\Name{Pound}{R.V.},
\REVIEW{\rmp}{71}{1999}{S54}.

\bibitem[\protect\citeauthoryear{Reinhold}{2006}]{Reinhold06}
\Name{Reinhold}{E.} \etal,
\REVIEW{\prl}{96}{2006}{151101}.

\bibitem[\protect\citeauthoryear{Reynaud}{2008}]{Reynaud08}
\Name{Reynaud}{S.} \etal,
\REVIEW{\pr}{D77}{2008}{122003}.

\bibitem[\protect\citeauthoryear{Reynaud}{2009}]{Reynaud09}
\Name{Reynaud}{S.} and \Name{Jaekel}{M.-T.}, to appear in
\BOOKed{International School of Physics Enrico Fermi on Atom Optics and Space Physics}
{}{2009}{Societa Italiana di Fisica}{}
[\arxiv{arXiv:0801.3407}].

\bibitem[\protect\citeauthoryear{Rosenband}{2008}]{Rosenband08}
\Name{Rosenband}{T.} \etal,
\REVIEW{Science}{319}{2008}{1808}.

\bibitem[\protect\citeauthoryear{Saathoff}{2003}]{Saathoff03}
\Name{Saathoff}{G.} \etal,
\REVIEW{\prl}{91}{2003}{190403}.

\bibitem[\protect\citeauthoryear{Salomon}{2001}]{Salomon01}
\Name{Salomon}{C.} \etal,
\REVIEW{C. R. Acad. Sci.}{IV-2}{2001}{1313}.

\bibitem[\protect\citeauthoryear{Salomon}{2007}]{Salomon07}
\Name{Salomon}{C.}, \Name{Cacciapuoti}{L.} and \Name{Dimarcq}{N.},
\REVIEW{\ijmp}{D16}{2007}{2511}.

\bibitem[\protect\citeauthoryear{Schlamminger}{2008}]{Schlamminger08}
\Name{Schlamminger}{S.} \etal,
\REVIEW{\prl}{100}{2008}{041101}.

\bibitem[\protect\citeauthoryear{Shapiro}{1999}]{Shapiro99}
\Name{Shapiro}{I.I.},
\REVIEW{\rmp}{71}{1999}{S41}.

\bibitem[\protect\citeauthoryear{Shapiro}{2004}]{Shapiro04}
\Name{Shapiro}{S.S.} \etal,
\REVIEW{\prl}{92}{2004}{121101}.

\bibitem[\protect\citeauthoryear{Schiller}{2009}]{Schiller09}
\Name{Schiller}{S.} \etal,
\REVIEW{\ea}{23}{2009}{573}.

\bibitem[\protect\citeauthoryear{Touboul}{2001}]{Touboul01}
\Name{Touboul}{P.} \etal,
\REVIEW{C. R. Acad. Sci.}{IV-2}{2001}{1271}.

\bibitem[\protect\citeauthoryear{Turyshev}{2006}]{Turyshev06}
\Name{Turyshev}{S.G.}, \Name{Nieto}{M.M.} and \Name{Anderson}{J.D.},
\REVIEW{EAS Publ.Ser.}{20}{2006}{243};
\Name{Turyshev}{S.G.}, \Name{Toth}{V.T.}, \Name{Kellogg}{L.R.} \etal,
\REVIEW{\ijmp}{D15}{2006}{1};

\bibitem[\protect\citeauthoryear{Uzan}{2003}]{Uzan03}
\Name{Uzan}{J.-P.},
\REVIEW{\rmp}{75}{2003}{403}.

\bibitem[\protect\citeauthoryear{Vessot}{1980}]{Vessot80}
\Name{Vessot}{R.F.C.} \etal,
\REVIEW{\prl}{45}{1980}{2081}.

\bibitem[\protect\citeauthoryear{Will}{2001}]{Will01}
\Name{Will}{C.M.}, \REVIEW{Living Rev. Rel.}{}{2001}{4}
\url{http://www.livingreviews.org/lrr-2001-4}.

\bibitem[\protect\citeauthoryear{Williams}{1996}]{Williams96}
\Name{Williams}{J.G.}, \Name{Newhall}{X.X.} and \Name{Dickey}{J.O.},
\REVIEW{\pr}{D53}{1996}{6730}.

\bibitem[\protect\citeauthoryear{Williams}{2004}]{Williams04}
\Name{Williams}{J.G.}, \Name{Turyshev}{S.G.} and \Name{Boggs}{D.H.},
\REVIEW{\prl}{93}{2004}{261101}.

\bibitem[\protect\citeauthoryear{Wolf}{1997}]{Wolf97}
\Name{Wolf}{P.} and \Name{Petit}{G.},
\REVIEW{\pr}{A56}{1997}{4405}.

\bibitem[\protect\citeauthoryear{Wolf}{2009}]{Wolf09}
\Name{Wolf}{P.} \etal,
\REVIEW{\ea}{23}{2009}{651}.

\end{thebibliography}
}
\end{document}